\documentclass[unnumbib]{comnet}

\usepackage{graphicx}
\usepackage{color}
\usepackage{amsmath}
\usepackage{amsfonts}
\usepackage{amssymb}
\usepackage{dcolumn}
\usepackage{booktabs}

\usepackage[toc,page]{appendix}

\usepackage[final]{pdfpages}

\begin{document}

\title{Smart Rewiring for Network Robustness}
\shorttitle{Smart Rewiring for Network Robustness} 
\shortauthorlist{Louzada et al.} 
\author{
  \name{V. H. P. Louzada}
  \address{Computational Physics, IfB, ETH Zurich, Wolfgang-Pauli-Strasse 27, Zurich, Switzerland \email{$^*$Corresponding author: louzada@ethz.ch}}
  \name{F. Daolio}
  \address{Faculty of Business and Economics, University of Lausanne, Lausanne, Switzerland}
  \name{H. J. Herrmann}
  \address{Computational Physics, IfB, ETH Zurich, Wolfgang-Pauli-Strasse 27, Zurich, Switzerland \\ Departamento de F\'isica, Universidade Federal do Cear\'a, 60451-970 Fortaleza, Cear\'a, Brazil}
  \and
  \name{M. Tomassini}
  \address{Faculty of Business and Economics, University of Lausanne, Lausanne, Switzerland}
}
\maketitle

\begin{abstract}
{While new forms of attacks are developed every day to compromise essential infrastructures, service providers are also expected to develop strategies to mitigate the risk of extreme failures. In this context, tools of network science have been used to evaluate network robustness and propose resilient topologies against attacks. We present here a new rewiring method to modify the network topology improving its robustness, based on the evolution of the network largest component during a sequence of targeted attacks. In comparison to previous strategies, our method lowers by several orders of magnitude the computational effort necessary to improve robustness. Our rewiring also drives the formation of layers of nodes with similar degree while keeping a highly modular structure. This ``modular onion-like structure'' is a particular class of the onion-like structure previously described in the literature. We apply our rewiring strategy to an unweighted representation of the World Air-transportation network and show that an improvement of 30\% in its overall robustness can be achieved through smart swaps of around 9\% of its links.}
{network robustness, risk analysis}
\end{abstract}

\section{Introduction}
\label{sec:intro}

The construction of a robust infrastructure network represents a great challenge to our society. In order to guarantee a broad and efficient coverage of basic services such as water, electricity, and telecommunications, decision makers need to take into account the effects of a great number of threats to the correct functioning of the system~\cite{Albert2000,Helbing2009}. Targeted terrorist attacks or random extreme weather conditions impose a systemic risk of catastrophic failure that has to be mitigated. In this way, tools provided by network science have offered interesting insights on common features of robust networks or methods and strategies to protect infrastructures~\cite{Frank1970,Molloy2000,Paul2004a,Ash2007,Moreira2009,Rokneddin2012}.

Consider the construction of an air-transportation network as an example, a challenge currently faced by many developing nations~\cite{Economist2012,France242012}. The localization of the airports should be decided given a tight supply and demand rule in order to ensure their efficiency, but other factors should also be included in the planning, such as security measures or noise reduction~\cite{Swissinfo2012}. Besides that, the overall system robustness should be taken into account, as the transportation of goods and people cannot be entirely halted in case that some airports close. 

When designing a new network from scratch, decision makers have an excellent opportunity to warrant its future robustness against failures~\cite{Havlin2012}. However, most of the current infrastructure has been built in a non-supervised fashion, mostly through a preferential attachment mechanism, where highly connected nodes (e.g. airports, Internet Service Providers) have a higher probability of receiving a new link (e.g. flights, transmission cables)~\cite{Barabasi1999}. Inspired by this situation, we propose in this work a strategy to improve the robustness of a given network by a small number of interventions, which makes the method useful for real-time actions under budget constrains.

Simple modifications of the network topology, the connection pattern of nodes through links, have been shown to be an effective way to increase the robustness under node or link attacks~\cite{Jiang2011,Zeng2012,Peixoto2012,Hayes2012,Wang2013}. In particular, Schneider et. al~\cite{Schneider2011} showed that successive random rewirings (link swaps) create a robust network through the formation of an onion-like structure in which high-degree nodes compose a core with further interconnected layers of radially decreasing degrees. 

In this work we propose a \emph{smarter} rewiring that lowers by several orders of magnitude the computational effort necessary to improve robustness. Our method is consistently better than the random rewiring for a small number of swaps and yields the same level of robustness in the long term limit. An onion-like structure is also created, although a higher modularity and degree correlation is observed in comparison to networks created by random swaps. We apply our rewiring strategy to the World Air-transportation network and we show that an improvement of 30\% in its overall robustness can be achieved through smart swaps of around 9\% of its links.

\section{Model}
\label{}

In a complex network, nodes (representing power stations, airports, proteins, etc.) interact through links (cables, flights, molecular binding, etc.) resulting in complex behavior that describes technical and biological systems~\cite{Watts1999,Latora2001,Maslov2002,Dorogovtsev2003,Newman2003,Caldarelli2007,Mamede2012}. Particularly, complex networks provide significant insights into a system robustness, either in a static~\cite{Holme2002,Albert2004,Sydney2010} or dynamic~\cite{Boccaletti1997,Motter2002,Wang2002b,Raposo2003,Shargel2003,Louzada2012} context. We focus here on a generic approach to improve network robustness and  consider only the simple case of networks where all links have the same importance (unweighted) and no orientation (undirected). For illustration purposes, we explain our model and related concepts in the framework of the World Air-transportation network, a system of paramount importance to our globalized world and that has been the subject of a lot of research in the past years~\cite{Guimera2005,Amiel2005,Colizza2006}. 

An Air-transportation network is defined here as \emph{robust} when it allows a passenger to travel between most of the airports even considering the disruption of the service in the major connection hubs, i.e., the airports with largest number of flights. This feature is directly associated to the size of the largest connected component (LCC) of a network. In mathematical terms, robustness $R$ is described as,
\begin{align}
R = \frac{1}{N}\sum_{p=1}^{N}S(\frac{p}{N})\ ,   
\end{align}
where $N$ is the number of nodes/airports, $p$ is the number of airports removed from the network, and $S(q)$ is the size of the LCC after a fraction $q=p/N$ of nodes were removed, considering that all incoming links/flights were also removed from the network. The parameter $R$ is contained in the interval $1/N\le R < 1$, and is a measure for robustness: a small $R$ is associated to a fragile network and a larger $R$ to a robust one.

To focus on targeted attacks, the node removal starts first with the highly connected nodes, the network hubs, which intuitively have the largest impact on the size of the LCC. After removing the more connected node we update the degrees (number of connections) of every node, and remove the next largest network hub. This process is further repeated until the network completely collapses.

\begin{figure}[!h]
\centering
\includegraphics[width=0.45\columnwidth]{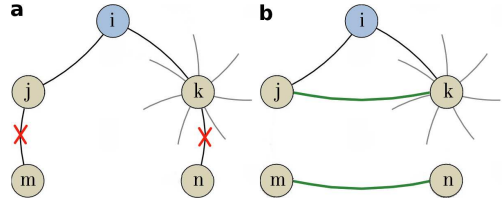}
\caption{\textbf{Smart rewiring for robustness improvement.} Diagrammatic representation of the smart rewiring. \textbf{a}, First steps of the smart rewiring: For a randomly selected node ($i$, blue), its lowest degree neighbor ($j$, brown) and highest degree neighbor ($k$, brown) are selected. In sequence, two neighbors of $j$ and $k$ are randomly selected ($m$ and $n$, both brown), and links to them ($e_{jm}$ and $e_{kn}$) are removed (red $X$). \textbf{b}, Last step of the smart rewiring in which links $e_{jk}$ and $e_{mn}$ (green) are added.} 
\label{fig::diagram}
\end{figure}

To improve robustness, one could simply add more flights between airports. In the limit, the network becomes fully connected: one airport disruption does not affect other destinations. But improving an airport flight capacity by adding redundancy might prove very impractical in the short term. In fact, numerous examples of infrastructure networks present this capacity constraint, such as adding new transmission lines to a power station or new traffic cables to an Internet Service Provider. Therefore, a rewiring strategy where links are only swapped, keeping the nodes' degree fixed, is more appropriate: we reroute flights from airports and create new connection possibilities, without considerably changing the airports' load.

Here we propose a novel rewiring strategy that improves network robustness by creating alternative connections between parts of the network that would otherwise be split upon the failure of
a hub. In our targeted attack scenario, we implicitly admit that the attacker perfectly knows the network degree sequence and thus can cause maximum damage. In the same way, we assume that the ``defender" knows that the attacker has this information and thus acts upon it through a smart rewiring defined as follows:
\begin{enumerate}
\item Select a node $i$ randomly with at least two neighbors with degree larger than one;
\item Select the lowest degree neighbor of $i$, the node $j$, and its highest degree neighbor, the node $k$;
\item Select randomly a neighbor $m$ of node $j$ and a neighbor $n$ of node $k$;
\item Repeat steps 1-3 until all nodes concerned are different from each other.
\item Remove links $e_{jm}$ and $e_{kn}$;
\item Create links $e_{jk}$ and $e_{mn}$.
\end{enumerate}
where $e_{ij}$ represents an undirected link between nodes $i$ and $j$. An illustration of this strategy is provided in Fig.~\ref{fig::diagram}. 
Swaps can provide positive or negative change in the robustness. Previous works have proposed different swap acceptance mechanisms~\cite{Herrmann2011,Louzada2012a} in order to increase robustness faster. To focus on the comparison of the random and smart strategies, we perform a simple greedy choice: at every step we compare the robustness before and after the swap, and consider it a successful step if the robustness has improved. If unsuccessful, the swap is reverted and another smart rewiring, or random rewiring for comparison, is tested. In what follows, we define $R_0$ as robustness of the network before any swap is executed, $R_1$ as robustness after one successful swap, and $R$ as its value after some steps are executed.

\begin{figure}[!h]
\centering
\includegraphics[width=0.55\columnwidth]{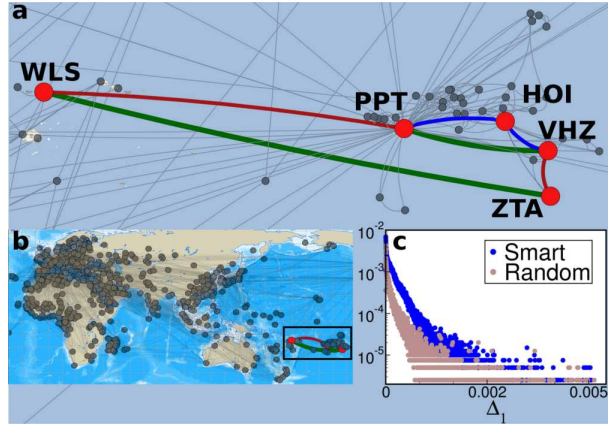}
\caption{\textbf{Proposed rerouting of flights for some airports in Oceania.} \textbf{a}, Example of the smart rewiring applied to the Hao Island Airport (HOI node), connected (in blue) to Faaa Airport (PPT), a regional hub, and to Vahitahi Airport (VHZ), a small airport. Connections Wallis Island (WLS) to Tureira Airport (ZTA) and PPT to VHZ are added (in green), while previous links from WLS to PPT and VHZ to ZTA are removed (in red). This simple swap increases the robustness of the World Air-transportation unweighted network by 1.85\%. \textbf{b}, Section of the World Air-transportation network showing the region in which airports in \textbf{a} are located. \textbf{c}, Effects of a single swap following the random and smart strategies on the overall robustness of a set of randomly generated Barab\'asi-Albert networks.} 
\label{fig::bestswap}
\end{figure}

\section{Onionlikeness}

The onion-like structure was first proposed by Schneider et. al~\cite{Schneider2011} as an emerging structure resulting from the random swap robustness optimization. To quantify this feature, we start by plotting the maximal number of nodes $S_k$ with degree $k$ that are connected through nodes with a degree smaller or equal to $k$. The onion-like structure presents more often paths between nodes of equal degree, which are not passing through nodes with higher degree, so a vertical positive shift in the $S_k$ curve is observed in comparison to a randomly generated BA network. Hence, a possible way to quantify this structure is through an \emph{onionlikeness} parameter $c$, the area below the $S_k$ curve,
\begin{equation}
c = \frac{1}{k^*}\sum_{k=1}^{k^*}\frac{S_k}{N_k} ,   
\end{equation}
where $k^*$ is the maximum degree among the nodes and $N_k$ is the number of nodes with degree $k$. In this formulation $1/k^* \leq c < 1$. At the lower bound, $c=1/k^*$, no special relation between a node degree and its neighbors' degrees are present. A regular lattice, for instance, where all nodes have the same degree, has $c=1/k^*$. The value of $c$ is close to the upper bound for networks with prominent onion structures, such as scale-free networks optimized for robustness.

\section{Results}

A swap keeps the number of links and nodes' degree unchanged, and is capable of changing the network robustness. A simple example is presented in Fig.~\ref{fig::bestswap}a-b for an unweighted representation of the World Air-transportation network. In this example, a single smart rewiring applied to an airport in Oceania is capable of improving the overall robustness by 1.85\%. If a swap is randomly executed, however, there is no guarantee that an improvement occurs, or that the magnitude of the improvement is satisfactory. Smart rewiring diminishes this problem as it presents a bias toward improvement. In a set of Barab\'asi-Albert (BA) networks, the distribution of the robustness improvement after one swap, $\Delta_1 = R_1 - R_0$, shows that significant changes of robustness are more common with our strategy (Fig.~\ref{fig::bestswap}c). Details regarding this and all other simulations are in Appendix~\ref{app:methods}.

\begin{figure}[!h]
\centering
\includegraphics[width=0.55\columnwidth]{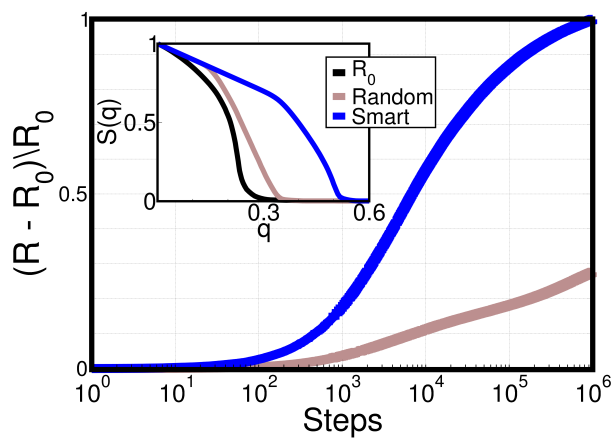}
\caption{\textbf{Fast improvement of network robustness for the smart rewiring strategy.} The smart rewiring allows a much faster improvement of $R$ in comparison to the random strategy. For $10^6$ steps, the inset shows the LCC during a sequence of targeted attacks. Data is an average of $100$ BA networks of $2005$ nodes (main plot) and $100$ BA networks of $1000$ nodes (inset).} 
\label{fig::robustness_evol}
\end{figure}

\begin{figure}[!h]
\centering
\includegraphics[width=\columnwidth]{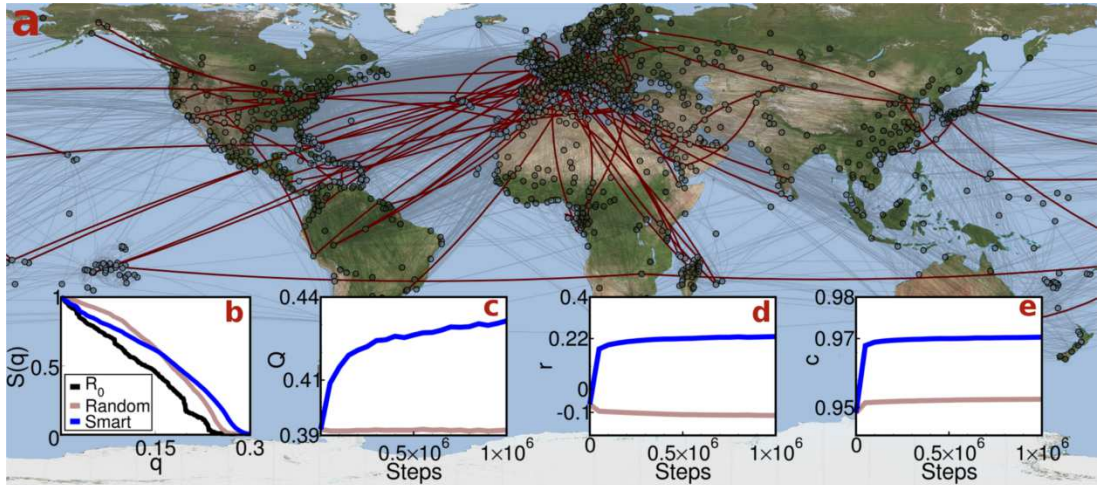}
\caption{\textbf{Robust Air transportation network}. \textbf{a}, The World Air-transportation network has its robustness improved by 4.82\% with swaps of 50 links (red) following the smart rewiring strategy. \textbf{b}, Size of the largest cluster for the World Air-transportation network through a sequence of targeted attacks before and after the application of the smart and random rewiring strategies. In this case, both strategies reach the same level of robustness (30\% of improvement), but while random rewiring changes 15.19\% $\pm$ $0.90\%$ of network links, smart rewiring changes only 9.24\% $\pm$ $0.53\%$. \textbf{c-e}, modularity ($Q$), assortativity ($r$), and onionlikeness ($c$) during the application of the random and smart rewiring strategies.} 
\label{fig::onion_airplane}
\end{figure}

If positive swaps are executed in sequence, a systematic increase in the network robustness is achieved. Schneider et al.~\cite{Schneider2011} showed an improvement of roughly 100\% in $R$ for a network of $N=1000$ after an extremely large number of swaps. Successive applications of the smart rewiring are much more efficient. We compare the evolution of $R$ in both methods starting from a set of BA networks in Fig.~\ref{fig::robustness_evol}, considering only the execution of sucessful swaps for both cases. While the smart rewiring doubles $R$ after roughly $10^6$ steps, random swaps are still at the level of $20\%$ improvement. The collapse of the LCC happens after a removal of 52\% of the nodes, a 50\% improvement over the random rewiring strategy (Inset of Fig.~\ref{fig::robustness_evol}). Tests for different network sizes show that the performance difference increases with network size (See Fig. S2 in Supplementary Material). In the limit of a large amount of swaps, random swaps can yield close to optimal robustness~\cite{Schneider2011}. Smart rewiring approaches the optimal robustness much faster and, consequently, both methods converge to the same level of robustness (See Fig. S3 in Supplementary Material). Successive swaps in the World Air-transportation network improve its robustness by $4.82\%$ with as few as $50$ positive swaps, $0.32\%$ of the total of links, as shown in Fig.~\ref{fig::onion_airplane}a. In this network, for a fixed level of robustness improvement ($30\%$), smart swaps affect only 9.24\% $\pm$ $0.53\%$ of the total of links, while random swaps have to change 15.19\% $\pm$ $0.90\%$ links (Fig.~\ref{fig::onion_airplane}b). 

Successive applications of the smart rewiring change drastically another characteristics of the network as well. Fig.~\ref{fig::onion_airplane}c shows the evolution of modularity~\cite{Blondel2008} ($Q$) during rewiring steps. The smart rewiring makes networks consistently more modular than random rewiring. This difference is a consequence of the intervention performed in the local connectivity by the smart rewiring, as our strategy deliberately creates triangles of connections. This structure reduces the importance of the hubs, which are now connected to leaves (nodes of low degree), and their removal does not have huge impact on global connectivity. These results are valid for different system sizes (See Fig. S2 in Supplementary Material).  

Despite the creation of connections between hubs and leaves, network assortativity~\cite{Newman2002} \emph{increases}, as the evolution of Newman's $r$ coefficient shows in Fig.~\ref{fig::onion_airplane}d. This result can be qualitatively understood considering the edges swapped. Before the rewiring, two edges contribute in a negative way to assortativity: $e_{jm}$ connects a leaf to an average degree node and $e_{kn}$ connects a hub to an average node. After the rewiring, one edge contributes negatively ($e_{jk}$ connects leaf to hub) and the other contributes positively ($e_{mn}$ connects average to average nodes). This effect is also persistent for different system sizes (See Fig. S2 in Supplementary Material) and considering assortativity through neighbor connectivity~\cite{Pastor-satorras2001} (See Fig. S1 in Supplementary Material). In comparison, both the original BA networks and networks optimized through random swaps are dissortative. 

Higher modularity and assortativity produced by the smart rewiring do not interfere with the formation of the onion-like structure, where layers of nodes of increasing degree hold the network robustness. Both strategies produce the onion-like structure but, by yielding a larger robustness, the onion structure is more prominent in the case of the smart rewiring (Fig.~\ref{fig::onion_airplane}e). Onionlikeness also remains larger for smart rewiring for different system sizes (See Fig. S2 in Supplementary Material).  

\section{Discussion}

Through a simple rewiring strategy we present here a method that improves drastically the network robustness while consuming little computational time. The proposed smart rewiring quickly increases robustness in comparison to a random choice of links. The high efficiency, together with the fact that only local knowledge of the two first neighbors of a given node is necessary, makes this strategy a potential tool for network designers and policy makers having the task of protecting our already built infrastructure against targeted attacks. As an example, simple interventions on the World Air-transportation network have been able to considerably improve its robustness. Our main analysis is performed on a set of randomly generated BA networks, which suggests that the same findings would apply to all real networks with a broad degree distribution since the smart swap is general and not limited to a particular network class.

Besides its simplicity, the smart strategy counterintuitively improves the maintenance of the largest cluster through a local division of the network: at each step five nodes previously connected are transformed into a triangle and a pair of nodes. This apparent division does not fully fragment the network, it only reduces the importance of the network hubs in keeping the global connectivity through the addition of links between nodes of average degree. These rewired links might eventually bridge different parts of the network after the hub failure. Moreover, smart rewiring creates also a highly modular and assortative topology while forming an onion-like structure. 

As modularity and assortativity differ radically from networks modified through random swaps, we define the structure of networks generated through successive applications of the smart rewiring as a \emph{modular onion} structure. This new topology gives rise to the question if further changes in the swap mechanism could create different structures. Following this, swap mechanisms could be designed to improve a certain desired feature, in the same way as the smart rewiring enhances modularity, while improving network robustness. As a method based on a simplified framework, another possible application of the current study is to adapt the strategy to real-time circumstances of an infrastructure network, such as flight capacity and climate conditions in the air transportation problem.

It is noteworthy that our model does not account for weights in the links, which would represent the number of passengers traveling between airports in a certain period of time. A rewiring method that takes advantage of this information, together with adaptations of the robustness concept, could have direct applications in the optimization of a real technical system.

\begin{appendix}
\section{Methods}
\label{app:methods}

The World Air-transportation network was retrieved from Amiel et. al~\cite{Amiel2005}. It contains data regarding only international airports and flights. The number of nodes/airports is $1326$, with $16001$ links, and average degree of $24.13$. Artificial networks considered in this work are all BA networks of average degree six.

In Fig.~\ref{fig::bestswap}, Panels \textbf{a} and \textbf{b} represent sections of the World Air-transportation. In particular, airports in Panel (a) are labeled according to their IATA code. Panel \textbf{c} is an average over $100$ BA networks of $1000$ nodes, the standard deviation of the points being smaller than the symbols.

In Fig.~\ref{fig::robustness_evol}, the main plot is an average over $100$ BA networks of $2005$ nodes. The inset is an average of $100$ BA networks of $1000$ nodes subjected to $10^5$ steps of rewiring (smart or random) in comparison to the original network. In both plots the thickness of the lines is bigger than the standard deviation.

The main plot of Fig.~\ref{fig::onion_airplane} contains the entire World Air-transportation network with rewired links in red and thicker. The location of some airports are slightly altered due to map projection distortions. Inset \textbf{b} contains data regarding the Air-transportation network before and after $10^5$ smart swaps, for which the smart rewiring curve is an average over $100$ different sequences of random swaps. Insets \textbf{c}-\textbf{e} are averages over 100 BA networks of $2005$ nodes. In all insets the thickness of the lines is bigger than the standard deviation.

\end{appendix}

\section*{Acknowledgment}
This work was supported by the Swiss National Science Foundation under contract 200021 126853, the CNPq, Conselho Nacional de Desenvolvimento Cient\'i­fico e Tecnol\'ogico - Brasil, the European Research Council under grant FP7-319968, and ETH Zurich Risk Center. Authors would like also to thank N. A. M. Ara\'ujo and K. J. Schrenk for the valuable discussions.

\clearpage

\begin{minipage}{\linewidth}
\begin{center}
\Large{\textbf{Supplementary Information}}
\vspace{5.0pt}

\end{center}

\end{minipage}

\renewcommand{\thefigure}{S\arabic{figure}}

\pagenumbering{gobble}
\setcounter{figure}{0}
\setcounter{section}{0}

\begin{center}
\section{Assortativity through neighbor connectivity}

\begin{figure}[!h]
\centering
\includegraphics[width=0.4\columnwidth]{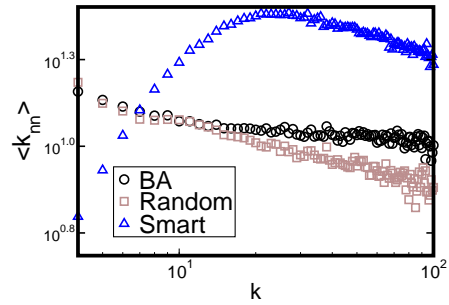} 
\caption{\textbf{Assortativity for different swap strategies}. Assortativity through neighbor connectivity~\cite{Pastor-satorras2001} for networks optimized using different strategies, and for BA networks in comparison. For each degree, $<k_{nn}>$ represents the average degree of the neighbors of nodes of degree $k$. Data is an average of $100$ networks of $2005$ nodes. The scattered values for large $k$ are due to statistical fluctuations.}
\label{fig::knn}
\end{figure}

\newpage
\section{System size effects}

\begin{figure}[!h]
\centering
\includegraphics[width=0.90\columnwidth]{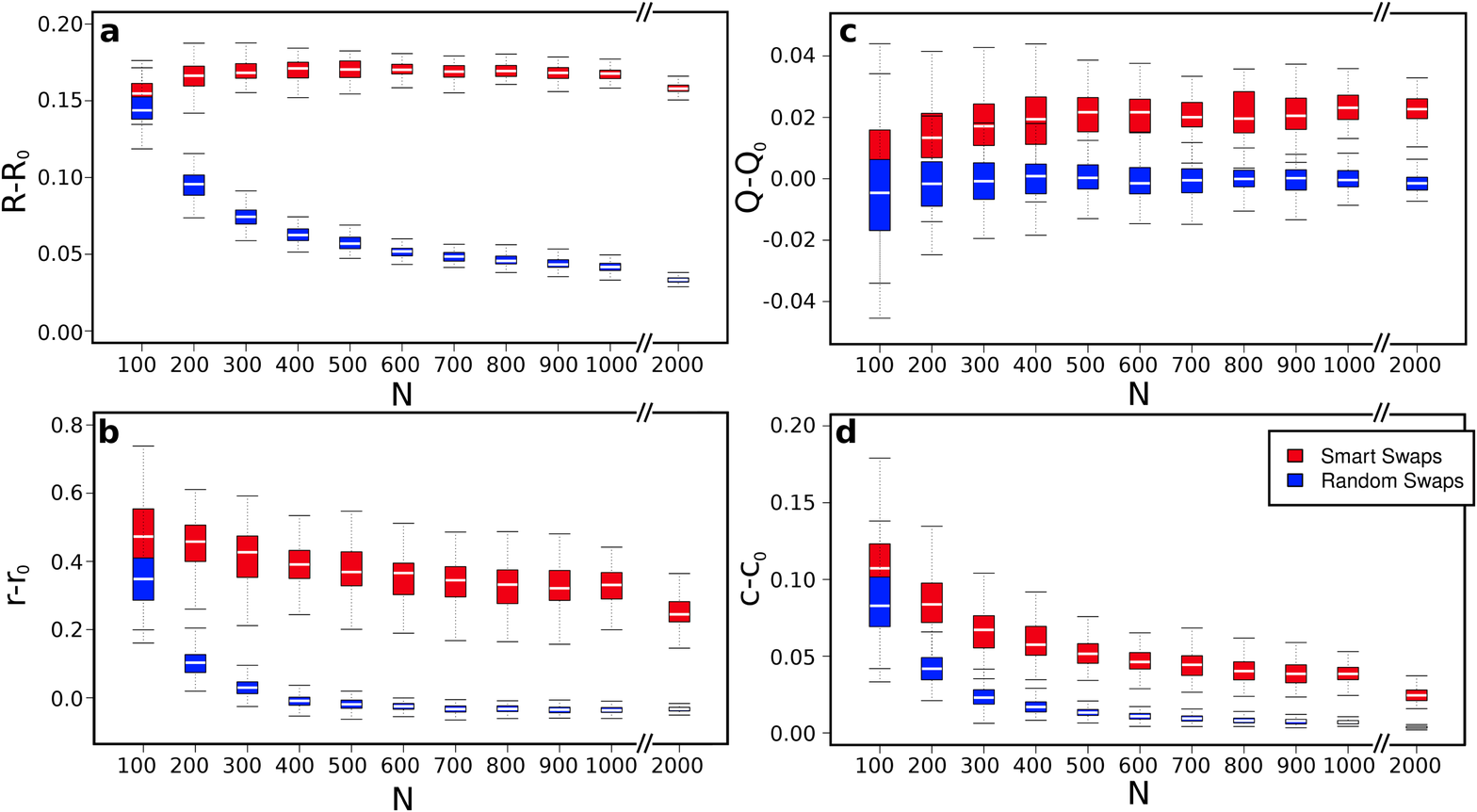} 
\caption{\textbf{Smart and random rewiring for networks of different sizes}. Effect of network size on Robustness (\textbf{a}), modularity (\textbf{b}), assortativity (\textbf{c}), and onionlikeness (\textbf{d}) for different system sizes. Each plot shows the difference between the quantity after $10^5$ steps ($R$, $Q$, $r$, and $c$) and its initial value ($R_0$, $Q_0$, $r_0$, and $c_0$). Box plots are used to represent the quantities computed for $100$ networks, according to: lower whisker for the lowest observation still within 1.5 IQR of the lower quartile, bottom of the box for the lower quartile, white trace for the median, top of the box for the upper quartile, and upper whisker for the highest value still within 1.5 IQR of the upper quartile.}
\label{fig::size}
\end{figure}

\section{Effect of the number of swaps}
\begin{figure}[!h]
\centering
\includegraphics[width=0.4\columnwidth]{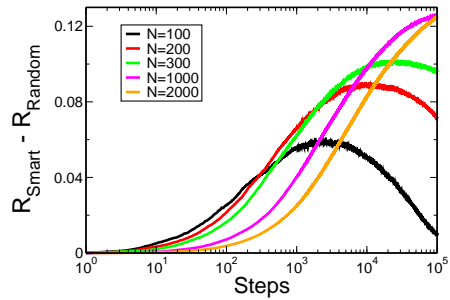} 
\caption{\textbf{Evolution of the difference between robustness for smart and random rewirings.} Comparing the difference between both methods, it is clear that for small networks a large number of swaps, either random or smart, lead to the same level of robustness. Each curve represents a system size. Data is an average of $100$ BA networks, with standard deviations smaller than curve thickness.}
\label{fig::nswap}
\end{figure}

\end{center}
\end{document}